\documentclass[journal,twoside,web]{ieeecolor}
\usepackage{jsen}
\usepackage{cite}
\usepackage{amsmath,amssymb,amsfonts}
\usepackage{algorithmic}
\usepackage{graphicx}
\usepackage{textcomp}
\usepackage{wrapfig}
\usepackage{stfloats}
\usepackage{booktabs}

\def\BibTeX{{\rm B\kern-.05em{\sc i\kern-.025em b}\kern-.08em
    T\kern-.1667em\lower.7ex\hbox{E}\kern-.125emX}}
\definecolor{abstractbg}{rgb}{0.89804,0.94510,0.83137}
\begin{document}
\title{Measurement of Material Volume Fractions in a Microwave Resonant Cavity Sensor Using Convolutional Neural Network}
\author{Mojtaba Joodaki, \IEEEmembership{Senior Member, IEEE} and Idriz Pelaj
\thanks{Manuscript received ???; revised ???; accepted
???. \textit{(Corresponding author: Mojtaba Joodaki.)}}
\thanks{Mojtaba Joodaki is with the School
of Computer Science and Engineering, Constructor University, 28759
Bremen, Germany (e-mail: mjoodaki@constructor.university).}
\thanks{Idriz Pelaj is with the School
of Computer Science and Engineering, Constructor University, 28759
Bremen, Germany (e-mail: idrizpelaj@gmail.com).}
}

\IEEEtitleabstractindextext{%
\fcolorbox{abstractbg}{abstractbg}{%
\begin{minipage}{\textwidth}%
\begin{wrapfigure}[18]{r}{4in}%
\includegraphics[width=3.8in]{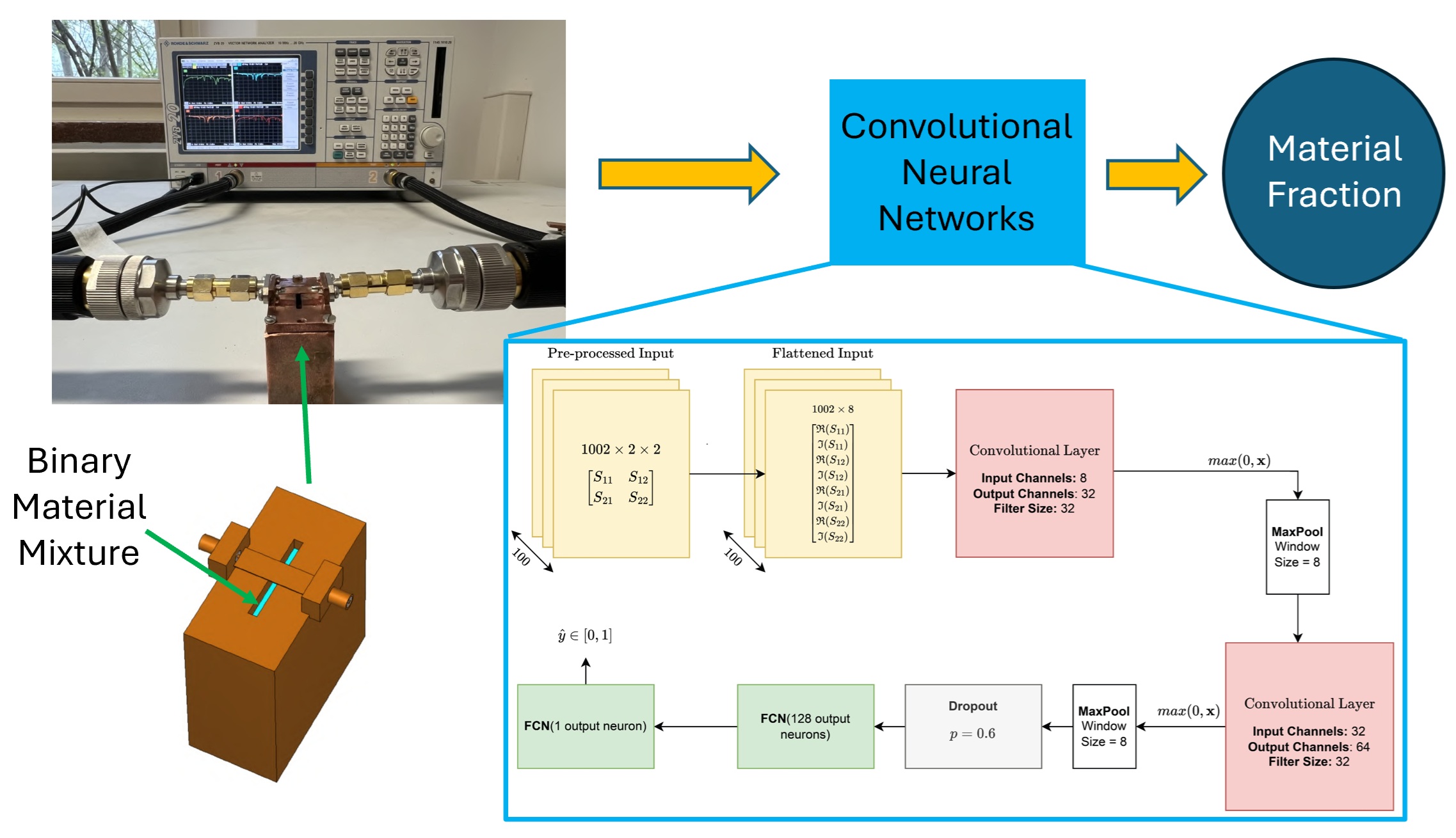}%
\end{wrapfigure}%
\begin{abstract}
A non-destructive, real-time method for estimating the volume fraction of a dielectric mixture inside a resonant cavity is presented. A convolutional neural network (CNN)-based approach is used to estimate the fractional composition of two-phase dielectric mixtures inside a resonant cavity using scattering parameter (S-parameter) measurements. A rectangular cavity sensor with a strip feed structure is characterized using a vector network analyzer (VNA) from 0.01--20~GHz. The CNN is trained using both simulated and experimentally measured S-parameters and achieves high predictive accuracy even without de-embedding or filtering, demonstrating robustness to measurement imperfections. The simulation results achieve a coefficient of determination ($R^2$)=0.99 using $k$-fold cross-validation, while the experimental model using raw data achieves an $R^2=0.94$ with a mean absolute error (MAE) below 6\%. Data augmentation further improves the accuracy of the experimental prediction to above $R^2=0.998$ (MAE$<$0.72\%). The proposed method enables rapid, non-destructive, accurate, low-cost, and real-time estimation of material fractions, illustrating strong potential for sensing applications in microwave material characterization.
\end{abstract}

\end{minipage}}}


\maketitle

\section{Introduction}
\label{sec:introduction}
\IEEEPARstart{A}{ccurate} determination of material composition is essential in numerous industries, including aerospace (fiber–resin ratios in composites), pharmaceutical manufacturing (monitor blending progress), automotive (biofuel blending), and archaeology (identification of historical artifacts) \cite{Rajan, Crouter, Vijayan, Creagh}. Traditional methods such as gravimetric analysis, chemical titration, or mass spectrometry are precise but often slow, destructive, time-consuming, require laboratory-grade equipment, and are not suitable for in-line monitoring \cite{skoog2022fundamentals}. In contrast, microwave sensing offers a rapid and non-destructive alternative for material characterization. Microwave sensors are increasingly used because they offer high accuracy, low maintenance requirements, and the capability for real-time sensing \cite{nelson0, Austin, joodaki2016a, joodaki2017, basseri2018}. Microwave resonant cavity techniques have long been used for non-destructive permittivity measurement \cite{Donovan, Nelson, Chen, mojgan, Peter, joo2024}. The resonance frequency and quality factor shift predictably with the effective permittivity of the material filling the cavity. However, analytical inversion from measured S-parameters to volume fractions in heterogeneous two-phase mixtures remains challenging due to nonlinear mixing laws and measurement uncertainties.

Recent advances in deep learning, particularly convolutional neural networks (CNNs), have shown remarkable success in inverse electromagnetic problems \cite{wei2019deep,li2020deep,Chrek, cas2025CNN, ali2025CNN}. CNNs have demonstrated robustness to noise, calibration imperfections, and coupling effects, making them suitable for RF measurement analysis. This paper presents a 1D-CNN that directly maps complex S-parameters of a rectangular cavity to the volume fraction of a binary dielectric mixture, eliminating the need for explicit resonance extraction or complex inversion algorithms.

In this paper, we propose and experimentally validate a CNN architecture capable of predicting fractional compositions of salt--sand mixtures inside a resonant cavity using measured S-parameters including the following major contributions:
\begin{itemize}
    \item A complete simulation-to-measurement pipeline using Bruggeman’s effective medium model and the required de-embedding process.
    \item Systematic study of preprocessing strategies (raw, augmented, filtered, de-embedded).
    \item Experimental validation with 21 salt–sand mixtures showing MAE $<$ 0.6\% and $R^2 > 0.999$.
\end{itemize}

The remainder of this paper is organized as follows. Section II explains the theoretical background including neural networks approaches used for S-parameter analysis, the proposed CNN and Bruggeman effective medium theory. The simulation and experimental setups and results are presented in Sections III and IV, respectively. Finally, the results and findings are discussed in Section V and the paper is concluded.

\section{Theoretical Background}
\subsection{Neural Networks Used for S-Parameter Analysis}
S-parameters describe the relationship between incident and reflected waves at microwave ports. For a two-port network,
\begin{equation}
\begin{bmatrix} b_1 \\ b_2 \end{bmatrix}=
\begin{bmatrix} S_{11} & S_{12} \\ S_{21} & S_{22} \end{bmatrix}
\begin{bmatrix} a_1 \\ a_2 \end{bmatrix}.
\end{equation}
where $a_1$ and $a_2$ are the incident power waves and $b_1$ and $b_2$ are the reflected waves. 

The complex-valued S-parameters encode resonance behavior, losses, and dielectric loading characteristics of the cavity under test. Neural networks have been widely employed to extract valuable information from measured S-parameters in microwave sensing applications. In a related study, Bartley \textit{et al.} \cite{nelson0} used a simple artificial neural network (ANN) to predict wheat moisture (10.6\%--19.2\% wet basis) from $S_{21}$ measurements at 10--18 GHz. The ANN had 16 inputs ($S_{21}$ amplitude/phase from eight frequencies), one hidden layer (15 neurons), and one output, trained via backpropagation on 179 samples with varying densities (0.72--0.88 g/cm$^{3}$) at 24°C. Split into train/test/production sets, it achieved 0.135\% MAE and $R^2$=0.99 compared with oven-dried references, demonstrating the usefulness of ANN for nondestructive moisture measurement in granules. 

Similarly, Chrek \textit{et al.} \cite{Chrek} proposed a deep neural network (DNN) model using a multilayer feedforward ANN to extract relative permittivity ($\epsilon_r$) and loss tangent ($\tan \delta$) of solids from 1--10 GHz S-parameters. A grounded coplanar waveguide fixture (GCPW fixture) with material under test (MUT) on top was used, with training data from full-wave simulations. Parametric optimization yielded seven hidden layers with doubled neurons, Xavier initialization \cite{Glorot}, Adam optimizer \cite{kingma2014}, and SELU (scaled exponential linear unit) activation. Validation on a known substrate showed a $\sim$1.2\% error. 

In another recent work, Khoshchehre \textit{et al.} \cite{ali2025CNN} integrated a dual-passband microstrip sensor (1.8--2.5 GHz \& 3.6--4.4 GHz) with a 1D-CNN to classify milk spoilage over 10 days at 21°C. $S_{21}$ spectra (101 points) of 10 samples (50 total, augmented to 250 with Gaussian noise $\sigma$=0.20--0.60 dB) showed shifts of 7.02 dB at 2.166 GHz. The CNN reached a 95.5\% training accuracy and 90\% validation accuracy, illustrating CNN effectiveness for non-invasive spectral analysis in food monitoring.

\subsection{Proposed CNN for Extracting the Material Volume Fractions from Measured S-parameters}
1D-CNNs are highly effective for spectral and time-series data \cite{kiranyaz2019cnn, Ige}. CNNs can extract localized features by applying learned convolutional filters across frequency-dependent S-parameter strings. In this work, each measurement consists of 1002 frequency points $(N_f=1002)$, where the real and imaginary parts of $S_{11}$, $S_{12}$, $S_{21}$, and $S_{22}$ are stacked into an 8-channel 1-dimensional (1D) tensor.

\begin{figure*}[h]
\centering
\includegraphics[width=0.8\textwidth]{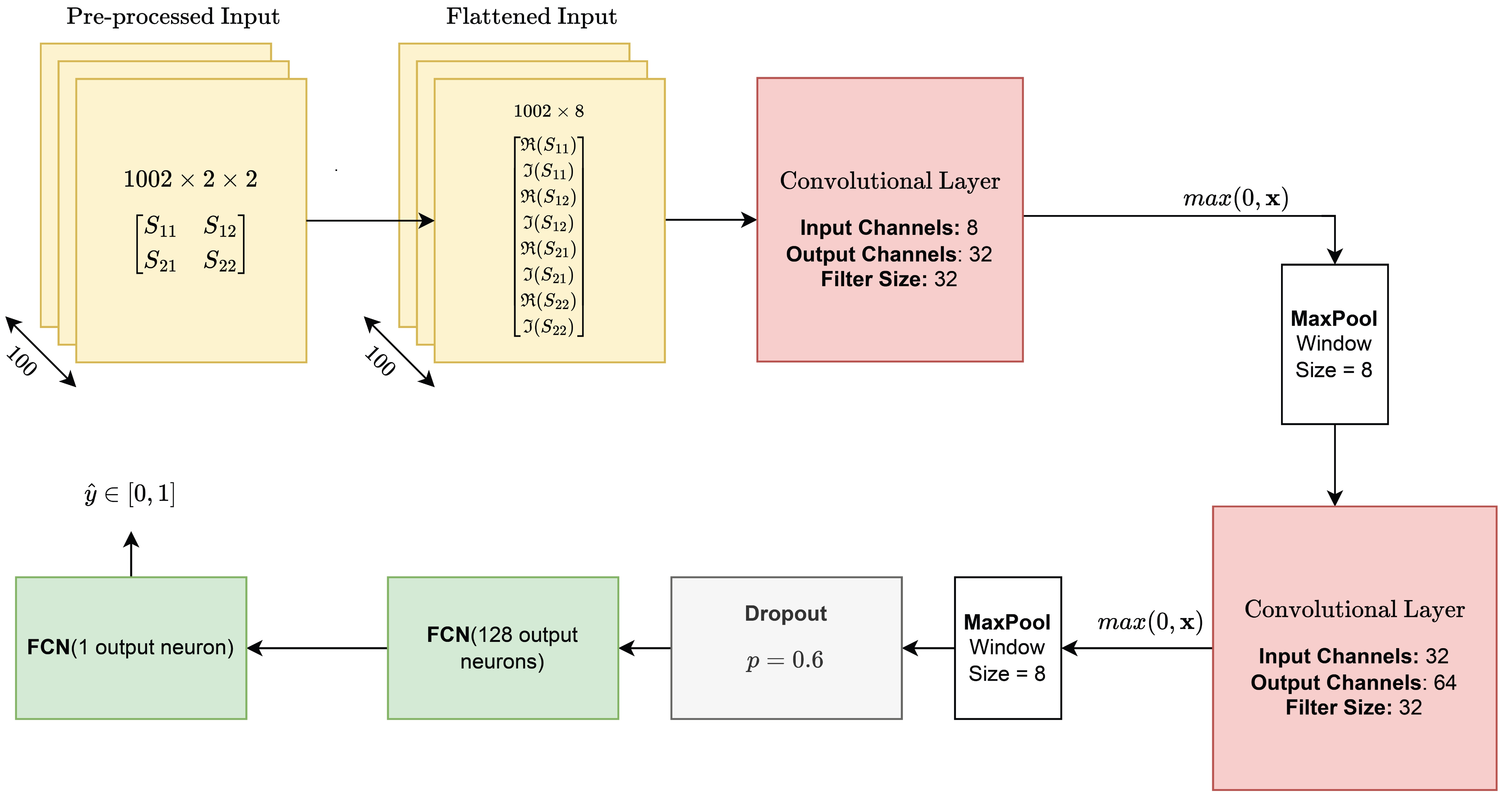}
\caption{The CNN architecture consists of two 1D convolutional layers, interleaved with max-pooling layers, and two fully connected layers at the end. The input data is a 1D array of size 1002 × 8 per sample.}
\label{fig:cnn_architecture}
\end{figure*}

As shown in Fig. 1, the following architecture was developed in Python using the PyTorch framework~\cite{paszke2019}:

\begin{itemize}
    \item \textbf{Input Layer:}  
    The input layer receives the complex S-parameters. The input tensor has a dimension of $1002 \times 8$, where 1002 corresponds to the number of frequency points and 8 represents the total number of input channels, obtained from the real and imaginary components of all four S-parameters.

    \item \textbf{First Convolutional Layer:}  
    The first convolutional layer applies 32 1D filters to the input data implemented using cross-correlation operations. A rectified linear unit (ReLU) activation is applied to the output of the convolutional layer. The ReLU function introduces nonlinearity into the network by allowing only positive values to pass through while setting negative values to zero, thereby enabling the model to learn complex nonlinear relationships~\cite{raman}.

    \item \textbf{Max-Pooling Layer:}  
    A max-pooling layer is employed to reduce the dimensionality of the feature maps while preserving the most salient features. The pooling operation selects the maximum value within each pooling window, allowing the network to focus on the most significant variations in the responses of the S-parameter.

    \item \textbf{Second Convolutional Layer:}  
    A second convolutional layer with an increased number of output channels is used to learn higher-level and more complex features from the input data. An additional ReLU activation is applied following the second convolutional layer to further enhance nonlinearity within the model.

    \item \textbf{Second Max-Pooling Layer:}  
    The convolutional feature-extraction stage concludes with another max-pooling layer, which performs additional downsampling and reduces the number of features that must be processed by the subsequent fully connected layers.

    \item \textbf{First Fully Connected Layer:}  
    The output of the convolutional stack is flattened into a one-dimensional vector and mapped to a fully connected layer containing 128 neurons.

    \item \textbf{Output Fully Connected Layer:}  
    The 128-neuron layer is subsequently mapped to a single output neuron, which represents the predicted material fraction.
    \end{itemize}

Since the predicted material fraction lies in the range of 0\% to 100\%, a sigmoid activation function is applied to the output neuron to map the prediction to the interval $[0,1]$. Only a single output is used, since the predicted value represents the fraction of a material constituent. The fraction of the second material can be directly obtained by
\begin{equation}
    \hat{y}_{\text{sand}} = 1 - \hat{y}.
\end{equation}

\subsection{Bruggeman Effective Medium Theory}
To estimate the effective dielectric constant of the sand–salt mixture used in the microwave resonant cavity simulations, the Bruggeman effective medium approximation is employed \cite{bruggeman1935}. This model treats the mixture as a symmetric, isotropic composite and accounts for the volume fractions of the constituent materials without assuming a distinct host–inclusion structure, making it particularly suitable for granular and powder-like mixtures such as sand and salt. While several alternative dielectric mixing models have been reported in the literature \cite{Nelson2}, including the Maxwell–Garnett, Lichtenecker, complex refractive index, and Landau–Lifshitz–Looyenga formulations, comparative studies have shown that the Bruggeman model provides reliable and physically consistent estimates for heterogeneous mixtures with comparable constituent fractions and randomly distributed particles \cite{Nelson1, Peltonen}. Given the moderate contrast between the dielectric properties of sand and salt and the focus of this work on relative changes in resonant behavior rather than absolute material characterization, the Bruggeman model offers an appropriate balance between accuracy, robustness, and computational simplicity. Therefore, it is used throughout this study.

For a two-phase mixture (host permittivity $\varepsilon_h$, inclusion permittivity $\varepsilon_i$ and volume fraction $f_b$), Bruggeman’s symmetric model gives \cite{bruggeman1935}:
\begin{equation}
f_b \frac{\varepsilon_i - \varepsilon_{\text{eff}}}{\varepsilon_i + 2\varepsilon_{\text{eff}}} + (1-f_b) \frac{\varepsilon_h - \varepsilon_{\text{eff}}}{\varepsilon_h + 2\varepsilon_{\text{eff}}} = 0.
\end{equation}
This equation is solved numerically for $\varepsilon_{\text{eff}}(f)$ and is used in all simulations.


 \section{Simulation Setup}
The microwave sensor, a rectangular resonant cavity with a central aperture and a feed microstrip line (as shown in Fig. 2), was modeled in CST Microwave Studio. The dielectric mixture inside the cavity is modeled using equation (3). One hundred simulations were performed for mixture fractions between 0--100\%. 
The S-parameters are obtained from the simulations and are used as input data for the CNN. However, some pre-processing is necessary. This includes de-embedding the parasitic effects from SMA connectors to aperture edges. In simulations, this is trivially done with the postprocessing function in CST Studio Suite, where we can de-embed the S-parameters up to the aperture edges from both ports uniformly. 

\begin{figure}[t]
\centering
\includegraphics[width=0.25\textwidth]{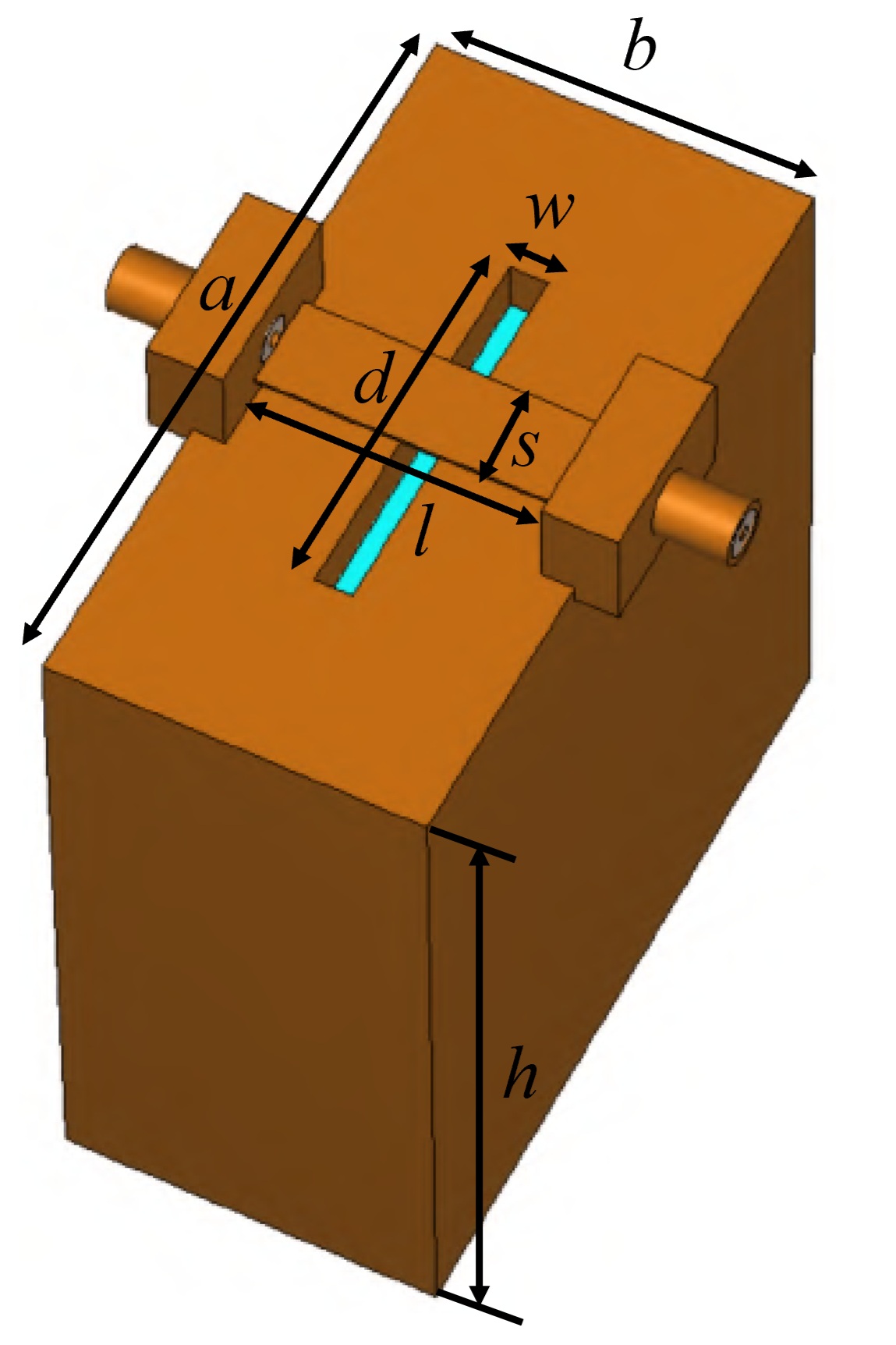}
\caption{The 3D sensor geometry used for the full-wave electromagnetic simulations in CST Studio Suite. The structure consists of a rectangular cavity with dimensions of 40 mm$\times$20 mm$\times$40 mm ($a \times b \times h$), featuring a centrally located aperture on the top surface measuring 20 mm$\times$2 mm ($d \times w$). A strip line of size 14 mm$\times$5 mm ($l \times s$) is positioned 1.9 mm above the cavity lid, oriented perpendicular to the longer side of the aperture~\cite{shour}}.
\label{fig:cnn_architecture}
\end{figure}

\begin{figure}[t]
\centering
\includegraphics[width=0.86\columnwidth]{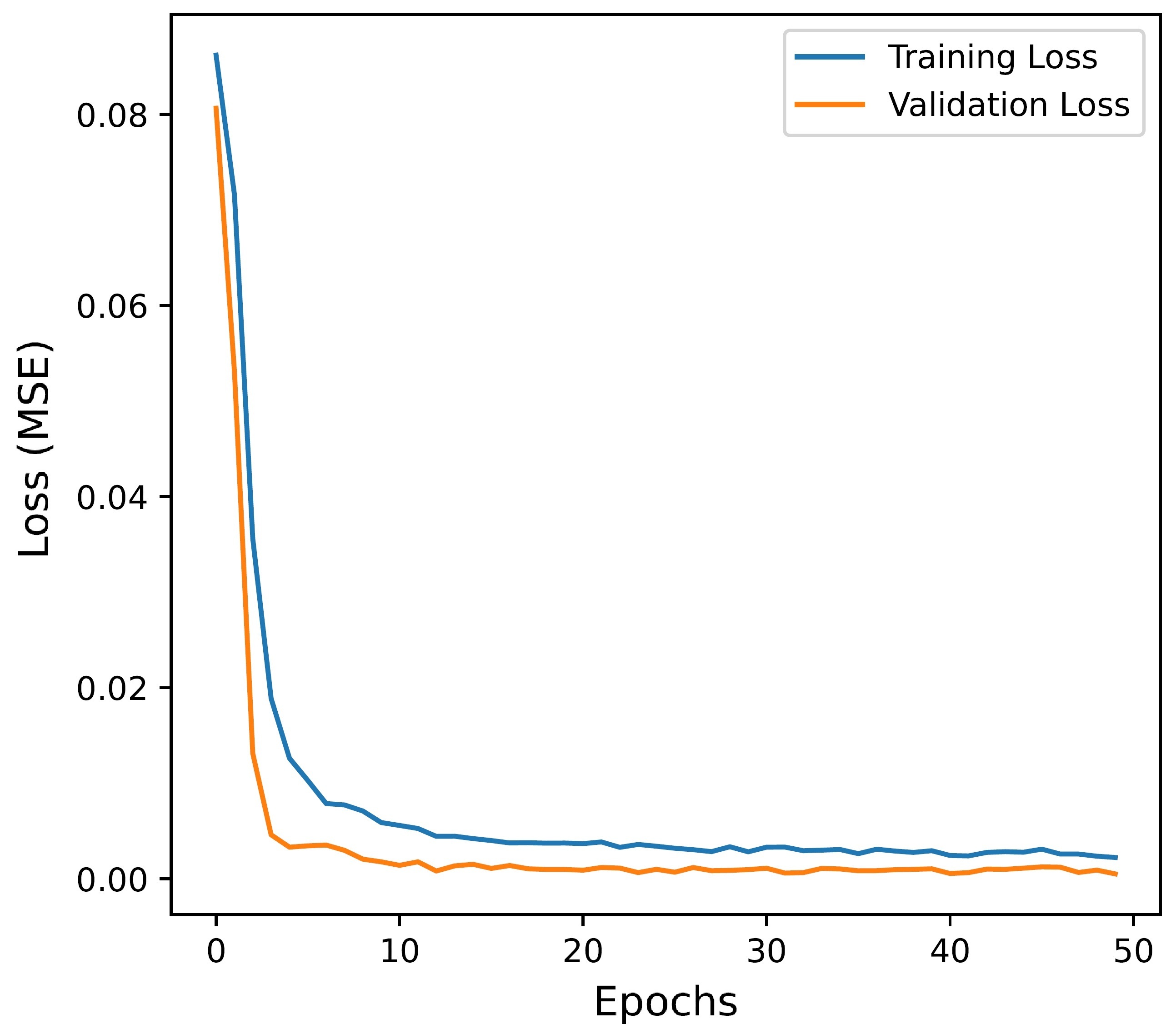}

(a)\\
\includegraphics[width=0.86\columnwidth]{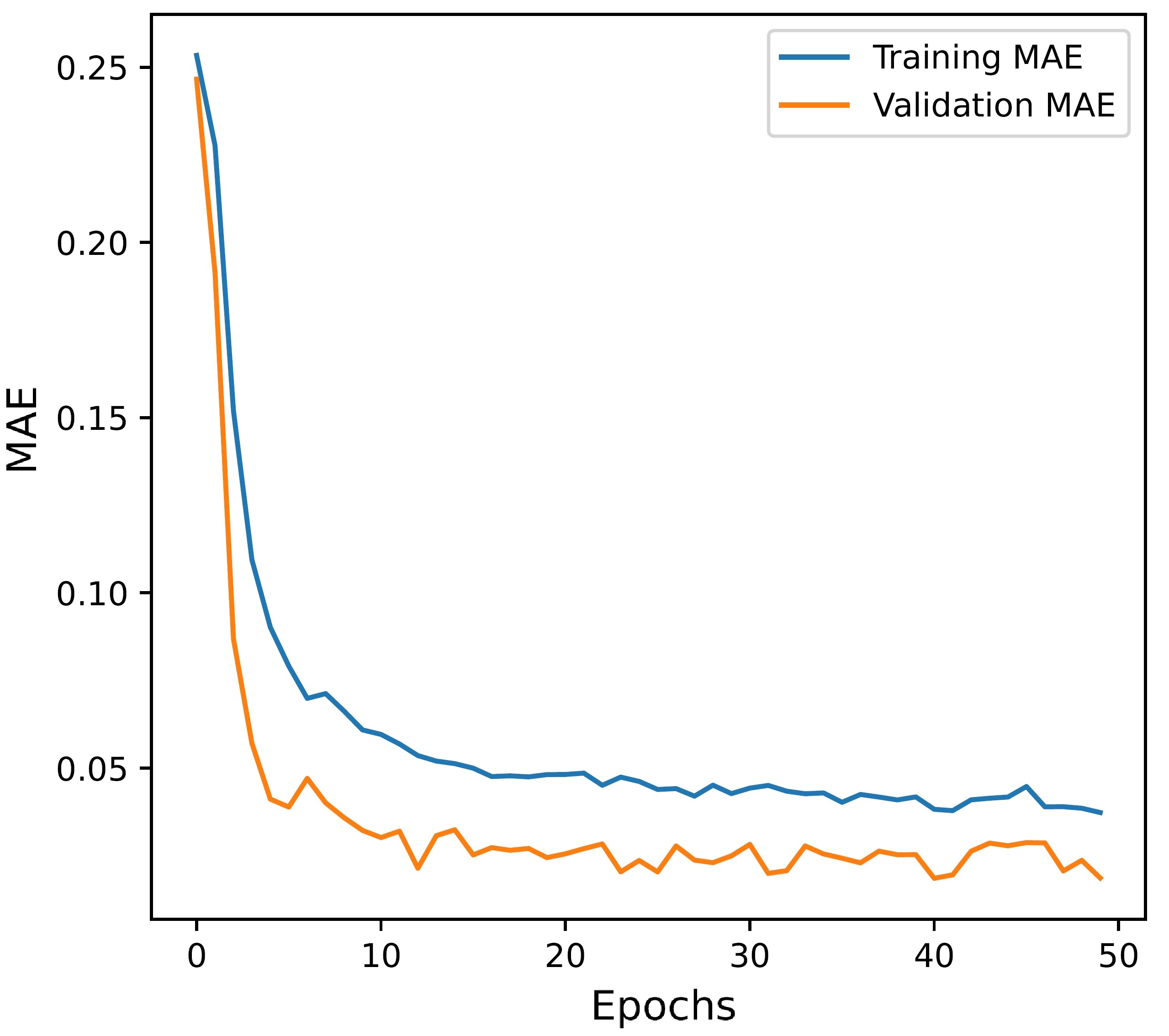}

(b)
\caption{The training and validation (a) losses and (b) MAEs of the CNN using simulated data averaged over five folds.}
\label{fig:cnn_architecture}
\end{figure}

The training and validation data are then split using $k$-fold cross-validation \cite{mclachlan2004}, with K = 5. In this work, we use the mean squared error (MSE) as the loss function, and the Adam optimizer \cite{kingma2014} is used to optimize the CNN weights. The training result using simulated data is presented in Fig. 3 which yields the following: 
\begin{itemize}
    \item MSE: $<10^{-4}$
    \item MAE: $<1\%$
    \item Coefficient of determination: $R^2=0.99$
\end{itemize}
These results demonstrate the feasibility of CNN-based fraction estimation under idealized conditions.


\section{Experimental Setup and Results}
\subsection{Experimental Setup}
A Rohde $\&$ Schwarz ZVB20 VNA was used to measure the S-parameters of 21 physical mixtures ranging from 0--100\% salt (intervals of 5\%). The VNA was set to sweep from 0.01 to 20 GHz. The measurement setup is shown in Fig. 4. 

A standard SOLT (Short–Open–Load–Through) calibration was performed up to the SMA connector reference planes. Subsequently, the de-embedding procedure described in~\cite{joo2019} was applied to shift the reference planes to the aperture edges. As explained in~\cite{joo2019}, three measurements are required for the de-embedding process: the S-parameters of the solid lid without an aperture (Fig. 5(a)), the S-parameters of the empty cavity with the perforated lid installed (Fig. 5(b)), and the S-parameters of the filled cavity with the perforated lid. De-embedding is the most time-consuming part of the measurement procedure. However, an important outcome of our experiments is that the de-embedding step may not be necessary, and skipping the de-embedding step does not affect the performance of the CNN considerably.

\begin{figure}[!h]
\centering
\includegraphics[width=\columnwidth]{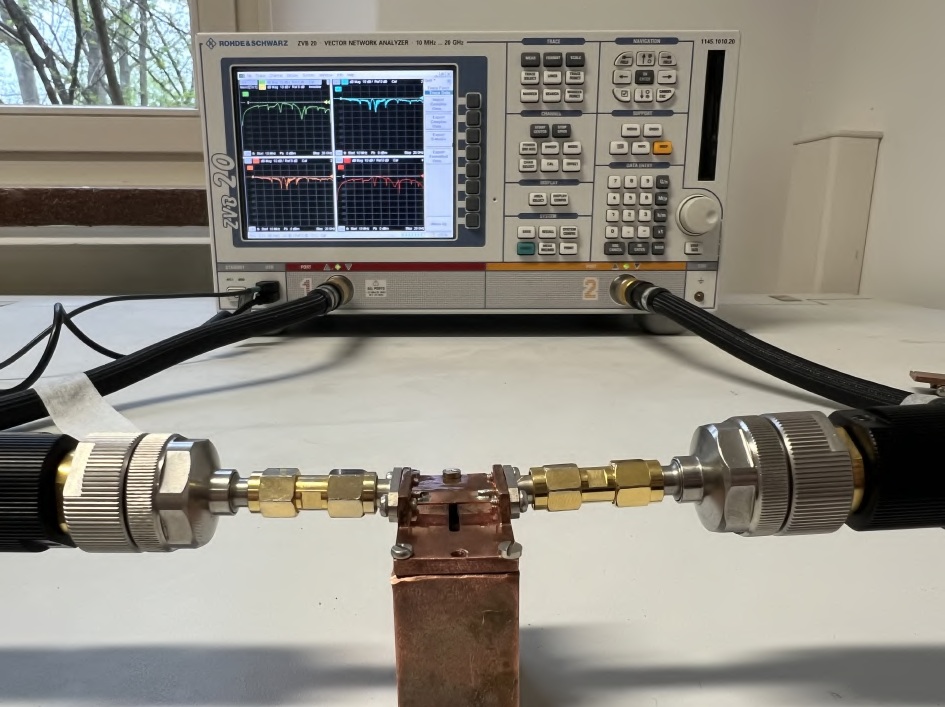}
\caption{Measurement setup for the binary material volume fraction sensor.}
\label{fig:cnn_architecture}
\end{figure}

\begin{figure}[!h]
\centering
\includegraphics[width=0.35\columnwidth]{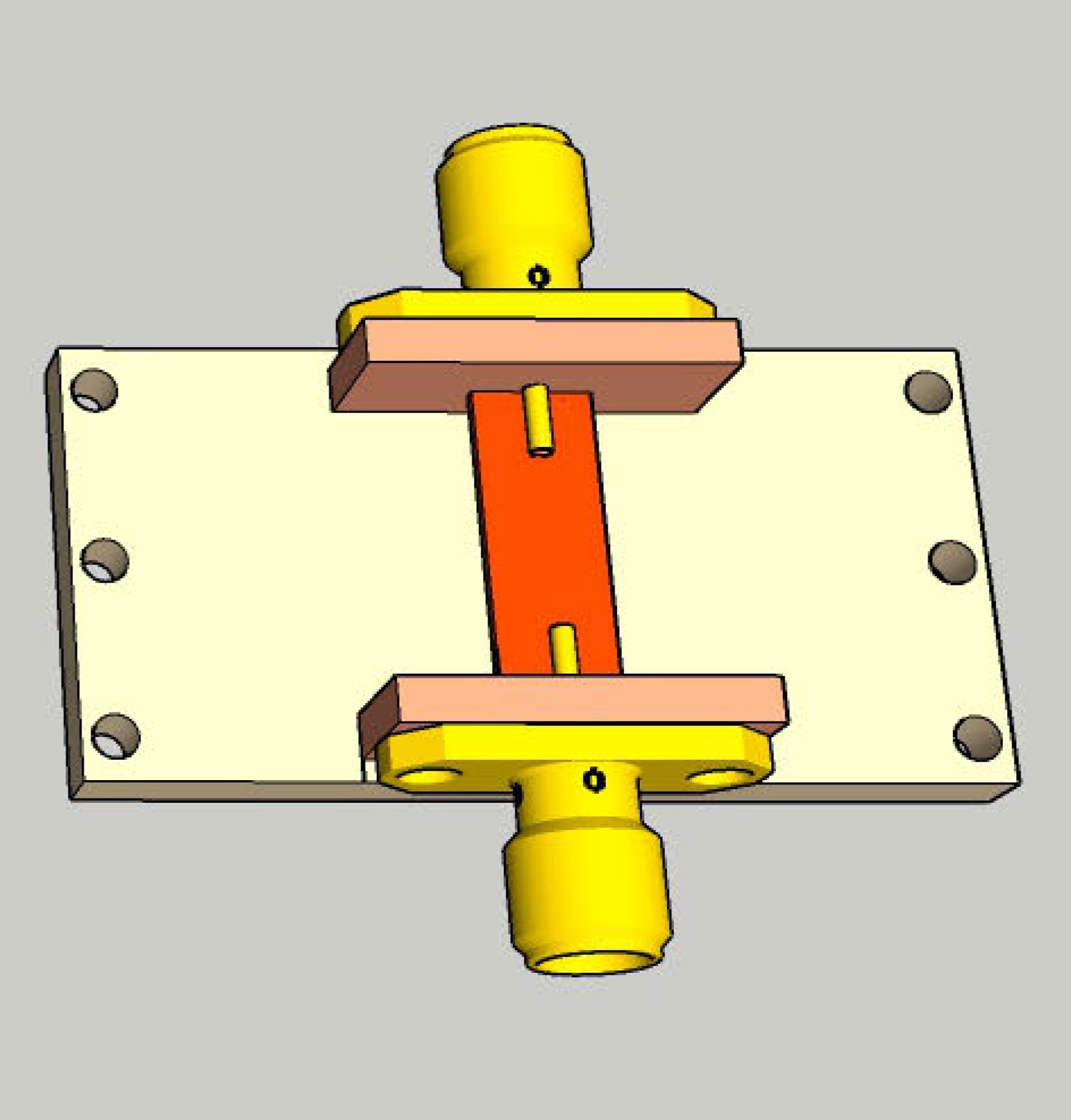}\\
(a) \\
\includegraphics[width=0.61\columnwidth]{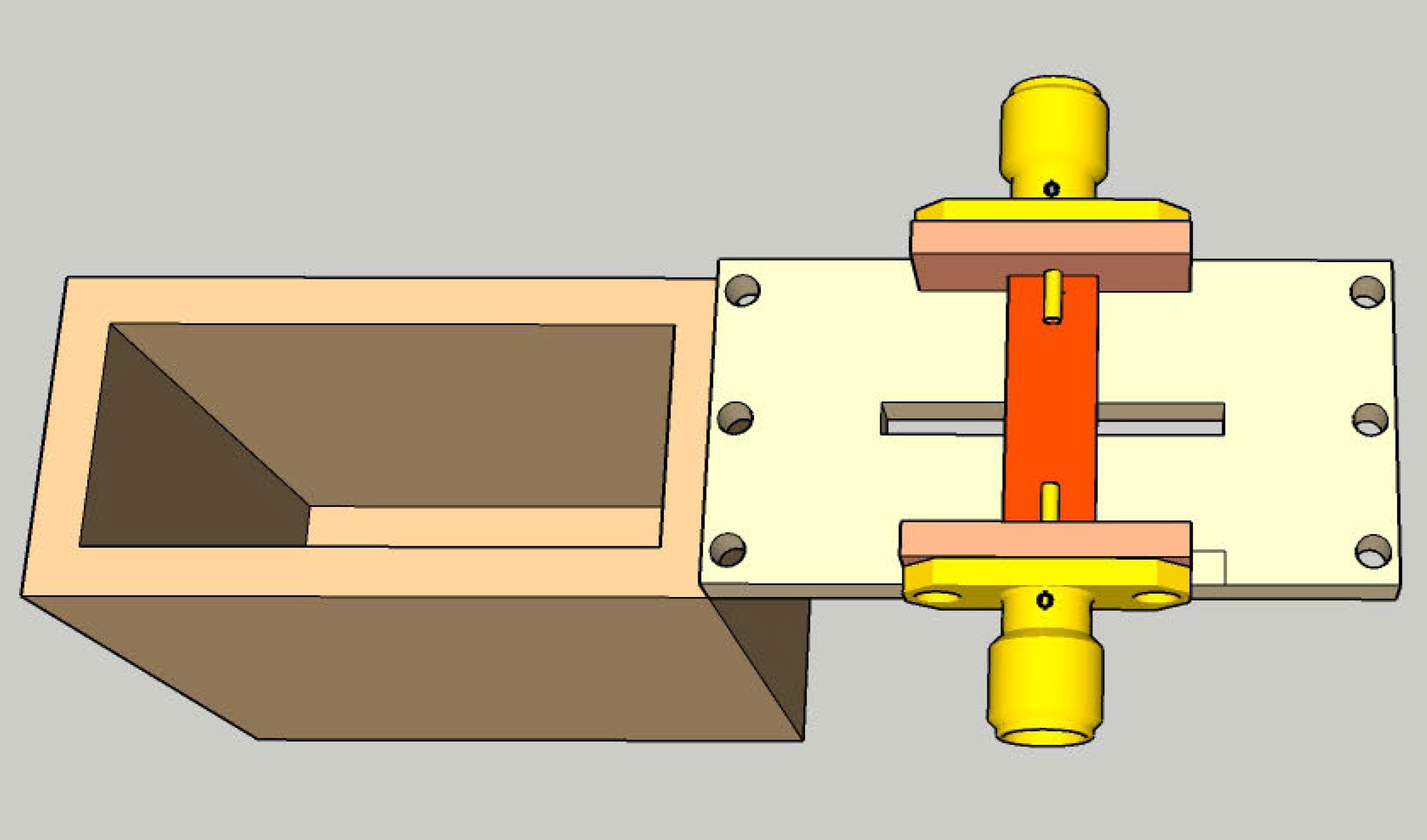}\\
(b)
\caption{(a) Solid lid (without aperture) used to de-embed the S-parameters up to the aperture edges. (b) Empty rectangular cavity with the perforated lid used for measuring the S-parameters of the empty sensor and the sensor loaded with the MUT.}
\label{fig:cnn_architecture}
\end{figure}

\subsection{Experimental Results}
In this work extensive experiments are conducted to examine various aspects of the proposed method, identify the optimal training strategy for learning S-parameter features, and analyze the impact of different parameters on the training process. Six different S-parameter datasets are utilized, each corresponding to the application of de-embedding, data augmentation, or filtering procedure, as described below:
\begin{itemize}
\item Raw S-parameters
\item Raw, augmented S-parameters
\item Raw, augmented, and filtered S-parameters
\item De-embedded S-parameters
\item De-embedded, augmented S-parameters
\item De-embedded, augmented, and filtered S-parameters
\end{itemize}

For instance, in the first scenario, only the raw measured S-parameters are used, with no de-embedding, data augmentation, or filtering applied. In contrast, the final scenario involves de-embedding the measured S-parameters to remove parasitic effects, augmenting the data to expand the dataset, and applying filtering to smooth the data and reduce noise. Data augmentation is performed via linear interpolation to generate intermediate mixture samples. The filtering is performed using a Savitzky–Golay filter with an MSE-based optimizer \cite{guest2013}.

We augment the S-parameters by a simple interpolation method, where we take the S-parameters corresponding to a percentage $x$ and $x$ + 5\% and interpolate four signals in between. This is done by a simple linear interpolation.Augmented data were excluded from validation to avoid artificial inflation of performance. Note that interpolation may introduce minor data leakage from training, which should be considered in real-world applications.

The raw measured S-parameters are presented in Fig. S1 in the supplemental material. In training with the raw S-parameters, we are extremely limited having only 21 samples to train and validate on. Our k-fold cross validation really struggles here, especially with such a small sample size, however, we find that the model behaves reasonably well with the limited data, with the validation loss (MSE) and mean absolute error (MAE) being relatively low. They are shown in Fig. 6. The training and validation losses of the CNN for the remaining scenarios are presented in Figs. S2–S6 in the supplemental material, respectively. 

Table~\ref{tab:perf} summarizes the CNN performance for different preprocessing strategies. Raw denotes the raw data from the VNA, De-emb denotes the de-embedded data, and the Aug and Filt suffixes denote whether the data was augmented or filtered, respectively. These validation metrics are taken from the best performing fold on each dataset.

\begin{figure*}[!h]
\centering
\includegraphics[width=0.8\columnwidth]{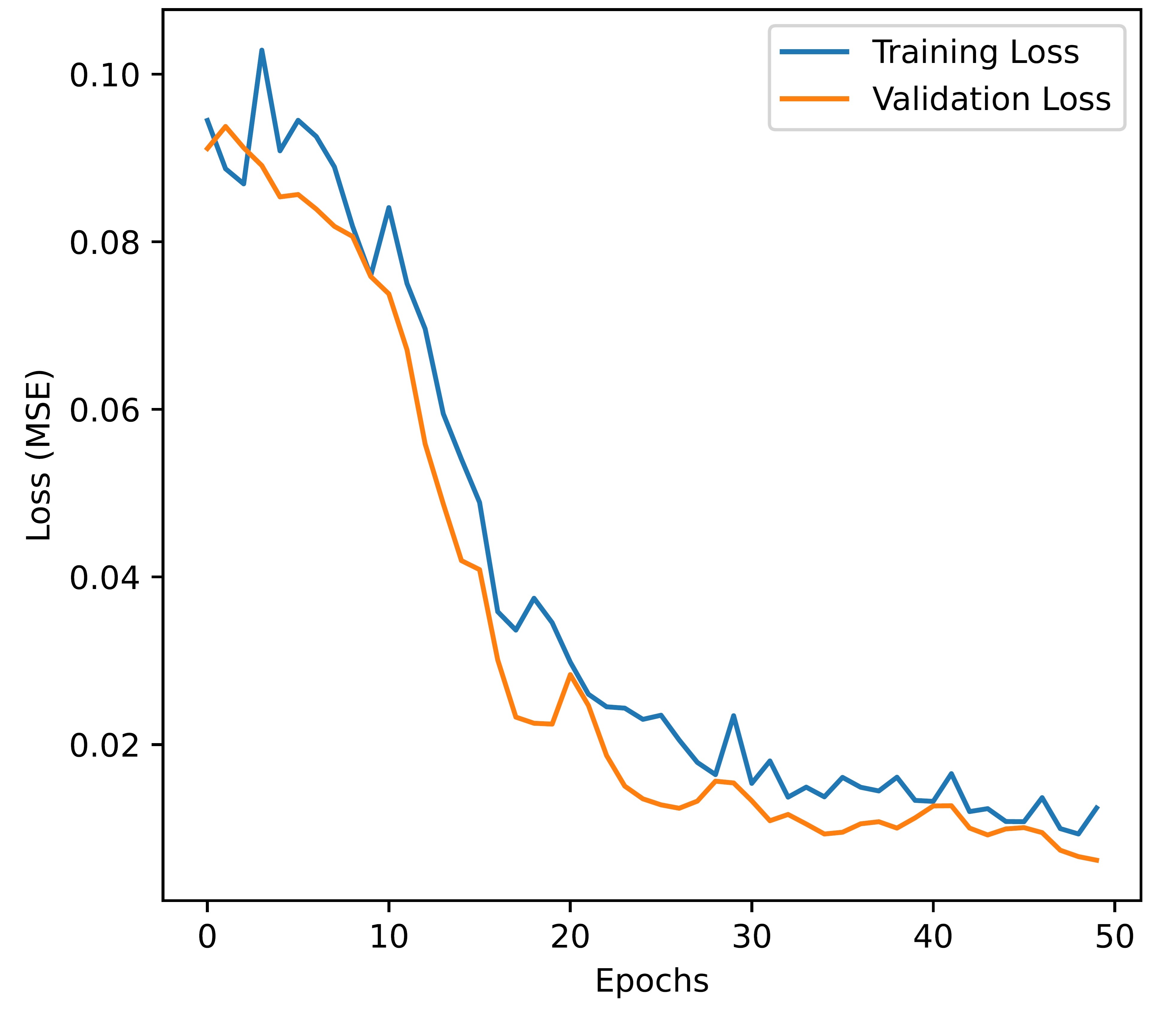}
\includegraphics[width=0.85\columnwidth]{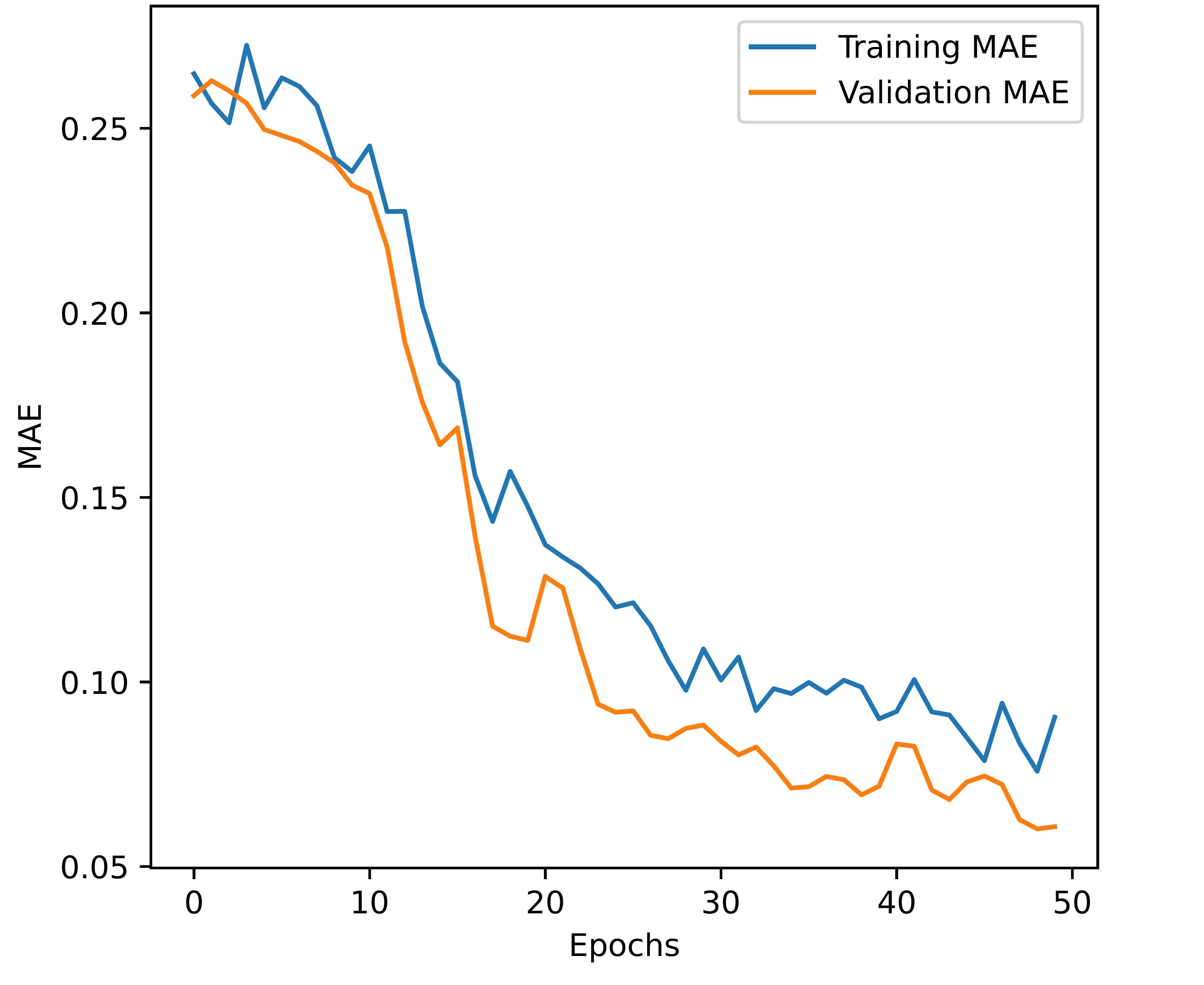}\\
(a) \;  \; \;  \; \;  \; \;  \; \;  \; \;  \; \;  \; \;  \; \;  \; \;  \; \;  \; \;  \;\;  \; \;  \; \;  \;(b) \\
\includegraphics[width=1.7\columnwidth]{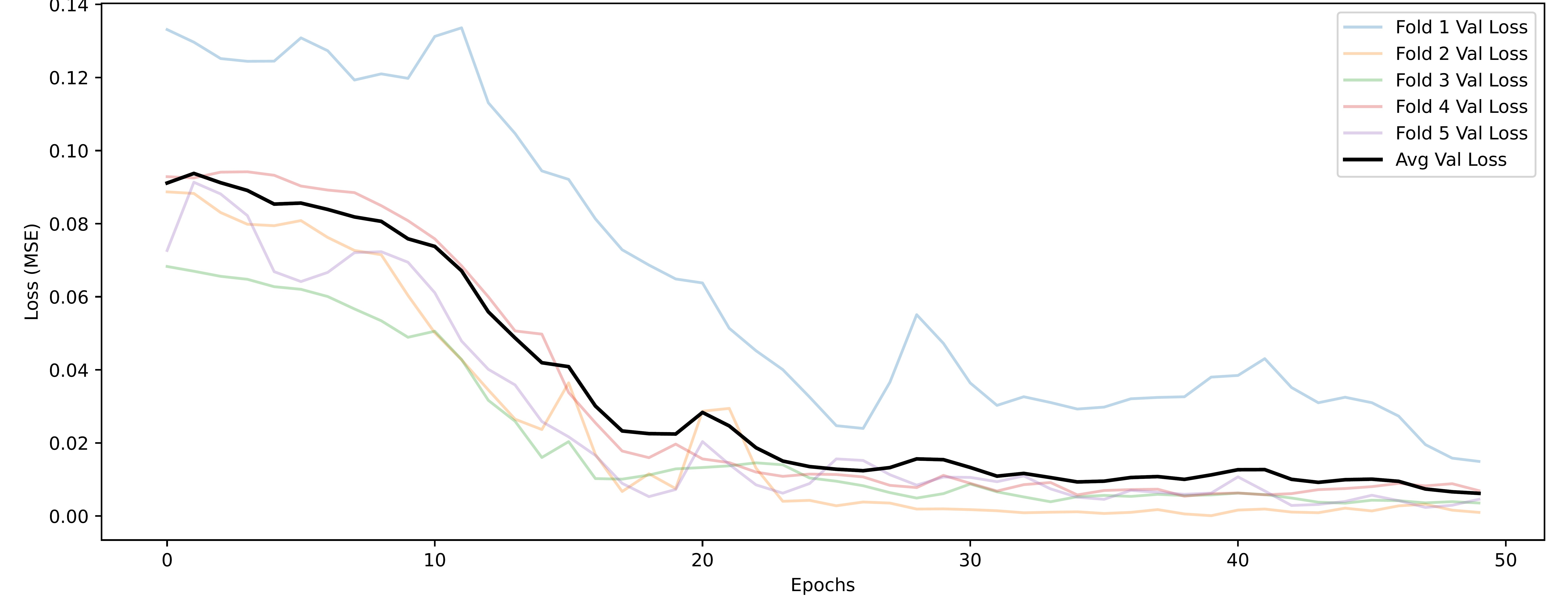}
\\(c)\\
\includegraphics[width=1.7\columnwidth]{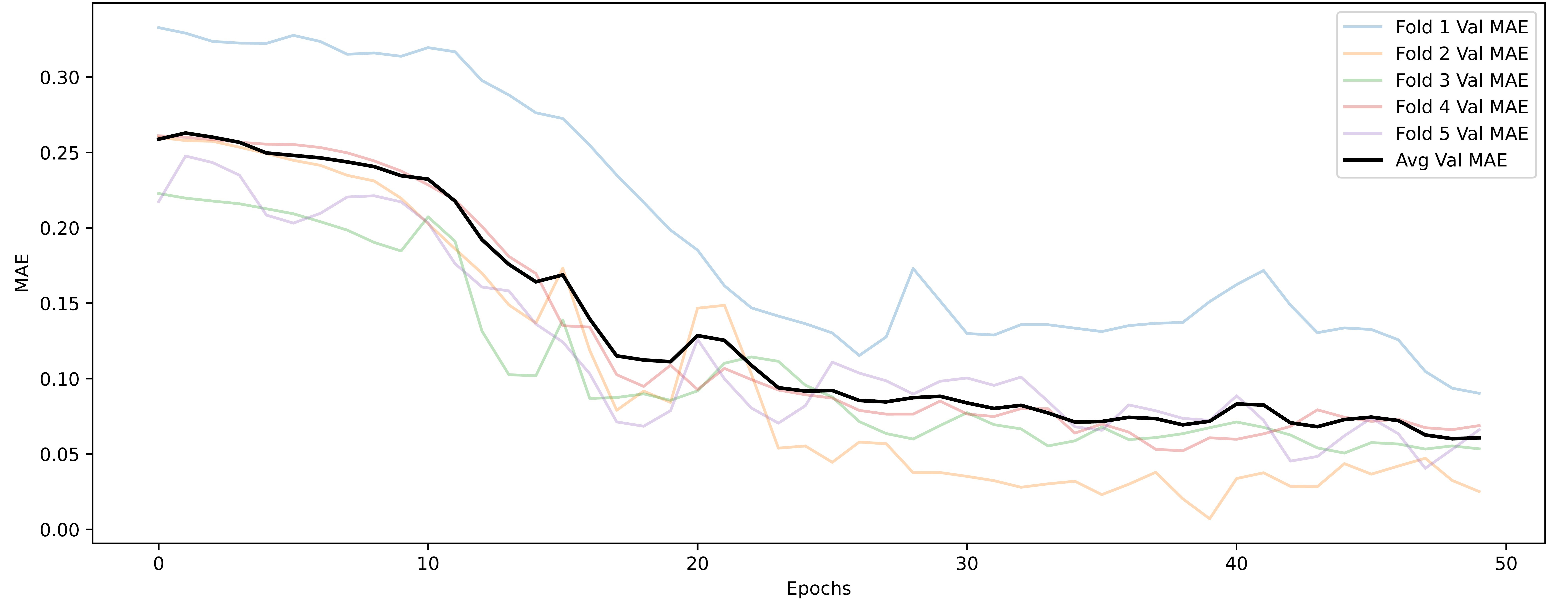}
\\(d)
\caption{The training and validation losses of the CNN using the raw S-parameters.(a) Losses averaged over 5 folds, (b) MAEs averaged over five folds, (c) losses for each fold, and (d) MAEs for each fold. }
\label{fig:cnn_architecture}
\end{figure*}

\begin{table}[h]
\centering
\caption{CNN Validation Performance for different Preprocessing Modes.}
\label{tab:perf}
\begin{tabular}{lccc}
\toprule
Dataset & MSE & MAE & $R^2$ \\
\midrule
Raw & $3.37\times10^{-3}$ & 0.0536 & 0.9419 \\
Raw+Aug & $7.1\times10^{-5}$ & 0.0072 & 0.9988 \\
Raw+Aug+Filt & $5.0\times10^{-5}$ & 0.0070 & 0.9991 \\
De-emb & $3.1\times10^{-3}$ & 0.0323 & 0.9637 \\
De-emb+Aug & $3.4\times10^{-5}$ & 0.0057 & 0.9994 \\
De-emb+Aug+Filt & $1.16\times10^{-4}$ & 0.0091 & 0.9983 \\
\bottomrule
\end{tabular}
\end{table}

\begin{table*}
\centering
\caption{Comparison of similar approaches for extracting material information from microwave S-parameters}
\label{tab:nn_comparison}
\begin{tabular}{p{2.2cm} p{3.3cm} p{3.3cm} p{3.6cm} p{3.8cm}}
\toprule
\textbf{Aspect} & \textbf{Bartley \textit{et al.} (1998)}~\cite{nelson0} & \textbf{Chrek \textit{et al.} (2022)}~\cite{Chrek} & \textbf{Khoshchehre \textit{et al.} (2025)}~\cite{ali2025CNN} & \textbf{This work} \\
\midrule

\textbf{Application} & Moisture content in wheat (10.6--19.2\,\% wet basis) & Retrieval of $\epsilon_r$ and $\tan\delta$ of solids & Milk spoilage classification (10 levels over 10 days) & Volume fraction of salt--sand mixtures in resonant cavity \\

\textbf{Sensor/Setup} & Free-space transmission (10.4\,cm wheat layer, horn antennas) & Grounded coplanar waveguide (GCPW), MUT on top & Dual-passband microstrip on RT/Duroid 4003 & Rectangular resonant cavity with microstrip feed \\

\textbf{Frequency Range} & 10--18\,GHz (8 points) & 1--10\,GHz (broadband) & 1.8--2.5 \& 3.6--4.4\,GHz (101 points) & 0.01--20\,GHz (1002 points) \\ 

\textbf{Neural Network Type}                 
                                 & Simple ANN (1 hidden layer, 15 neurons) 
                                 & Deep feedforward neural network (7 hidden layers, neurons doubled) 
                                 & 1D-CNN 
                                 & 1D-CNN (2 conv + 2 max-pool + FC layers) \\ 

\textbf{Inputs}                  & $S_{21}$ amplitude + phase (16 values) 
                                 & Full S-parameters (simulated) 
                                 & $S_{21}$ amplitude spectra (101 points) 
                                 & Complex $S_{11}$/$S_{12}$/$S_{21}$/$S_{22}$ (8 channels, 1002 points) \\ 

\textbf{Data Source}             & Experimental (179 wheat samples) 
                                 & Full-wave EM simulations 
                                 & Experimental (50 spectra, augmented to 250) 
                                 & Simulated (Bruggeman) + experimental (21 mixtures) \\ 

\textbf{Performance}             & MAE = 0.135\,\%, $R^2$ = 0.99 
                                 & $\sim$1.2\% error vs. reference 
                                 & 95.5\% training, 90\% validation accuracy 
                                 & Sim: $R^2$=0.99; Exp raw: $R^2$=0.94; Augmented: $R^2>$0.998, MAE$<$0.72\% \\ 

\textbf{Key Advantage}           & Density-independent moisture prediction 
                                 & Simulation-driven broadband dielectric retrieval 
                                 & Real-time non-invasive food quality monitoring 
                                 & Robust real-time fraction estimation even on raw/noisy data \\
\bottomrule
\end{tabular}
\end{table*}
\section{Discussion}
In the simulation, we can observe that the CNN is able to learn the features of the S-parameters and the relationship between the material fractions. This is crucial as it shows that the CNN not only generalizes to both augmented and non-augmented simulated data. 

For the raw S-parameters, we find that the CNN can learn the key features; however, it does not show the behavior of asymptotically converging even at 50 epochs. This is likely due to the fact that the data are very limited.  

In the data augmentation case, we see that the model performs much better, but this comes at the cost of relying on augmented data, which may not fully reflect real-world scenarios. This also shows that the model can learn the features of the S-parameters and the relationship to the material fractions without needing the de-embedding step, which could cost a lot of time in real-world applications.

There is minimal improvement from applying filtering to the raw, augmented data compared to not filtering it. This means that the filtering step is not necessary and that the CNN is able to learn the features of the S-parameters without needing additional filtering, once more proving the robustness of the CNN. 

Contrary to what one might expect, the de-embedding step seems to make the CNN perform worse. As shown in Fig. S4, the validation loss is higher than the training loss, and at 50 passes, it seems to be trending upward, oddly reflecting overfitting behavior. This may be an artifact from the de-embedding step, where floating point precision errors may be occurring.

Again, as is shown in Fig. S5, it is obvious that the data augmentation leads to a better performance than the raw de-embedded S-parameters. The validation and training losses are close to each other, and the oscillation observed in the case of the raw de-embedded S-parameters is no longer present.

Finally, as expected, the results presented in Fig. S6 show that filtering the de-embedded and augmented S-parameters will not help in improving the sensor performance.

To the best of our knowledge, this study represents the first reported use of a CNN to determine material volume fractions directly from broadband S-parameter measurements of a resonant cavity sensor. As no directly comparable microwave cavity-based systems employing deep learning for this specific task have been published so far, we benchmark our approach against related works that utilize neural networks for extracting material-related information from S-parameters in other microwave sensing configurations. A detailed performance comparison is provided in Table II.    

 \section*{Conclusion}
This work demonstrates that CNNs can accurately estimate fractional material compositions inside a resonant cavity using S-parameters. The method performs well with simulated and experimental data, achieving high accuracy with minimal preprocessing.

A key finding is that de-embedding and filtering---typically essential for accurate RF analysis---do not significantly improve CNN performance. The CNN learned robust features from raw S-parameters despite measurement imperfections. This is promising for real-time and field-deployable sensors where calibration time or fixture modeling is limited.

\section*{Acknowledgment}
The authors gratefully acknowledge Prof. Dr. Giuseppe Thadeu Freitas de Abreu of Constructor University, Bremen, Germany, and Mr. Roshan Nepal of the Department of Electrical and Computer Engineering, University of Waterloo, Canada, for their technical support and valuable contributions. The authors also thank all members of the MEND (Modern Electromagnetics and Nanoelectronic Devices) research group at Constructor University for their scientific and technical support.

\end{document}